\documentclass[pra,twocolumn,showpacs,preprintnumbers,amsmath,amssymb]{revtex4}

\usepackage{graphicx}
\usepackage{dcolumn}
\usepackage{bm}

\begin{document}


\title{Jarzynski Equality with Maxwell's Demon}
\author{Takahiro Sagawa$^1$}
\author{Masahito Ueda$^{1,2}$}
\affiliation{$^1$Department of Physics, Tokyo Institute of Technology,
2-12-1 Ookayama, Meguro-ku, Tokyo 152-8551, Japan \\
$^2$Macroscopic Quantum Control Project, ERATO, JST, 2-11-16 Yayoi, Bunkyo-ku, Tokyo 113-8656, Japan
}
\date{\today}

\begin{abstract}
We propose a new thermodynamic equality and several inequalities concerning the relationship between work and information for an isothermal process with Maxwell's demon.  Our approach is based on the formulation \`{a} la Jarzynski of the thermodynamic engine and on the quantum information-theoretic characterization of the demon. The lower bound of each inequality, which is expressed in terms of the information gain by the demon and the accuracy of the demon's measurement, gives the minimum work that can be performed on a single heat bath in an isothermal process. These results are independent of the state of the demon, be it in thermodynamic equilibrium or not.  
\end{abstract}


\maketitle

\section{Introduction}

Ever since the proposition of the ``demon'' by Maxwell \cite{paper1}, numerous studies have been conducted on the consistency between the role of the demon and the second law of thermodynamics \cite{paper2}. Bennett resolved the apparent contradiction by considering the logically irreversible initialization of the demon \cite{paper3}.  The key observation here is the so-called  Landauer principle \cite{paper4} which states that, in erasing one bit of information from the demon's memory, at least $k_{\rm B} T \ln 2$ of heat should, on average, be  dissipated into the environment with  the same amount of work being performed on the demon. Piechocinska has proved this principle without invoking the second law in an isothermal process \cite{paper5}. 

The essence of consistency between the role of the demon and the second law of thermodynamics can be illustrated by the setup of the Szilard engine \cite{paper6}. Suppose that the  entire state of the Szilard engine and the demon is initially in thermal equilibrium. The demon gains one bit of information on the state of the Szilard engine. The engine performs just $k_{\rm B} T \ln 2$ of work by using this information, before returning to the initial state. The demon then erases the obtained information from its memory. Consequently, the entire state returns to the initial equilibrium state. The sum of the work performed on the engine and the demon in a full cycle of the Szilard engine is non-negative according to the Landauer principle; thus the Szilard engine is consistent with the second law in this situation. However, the Landauer principle stated above tells us nothing if the demon is  far from equilibrium in the initial and/or final states.

Further discussions on Maxwell's demon involve quantum-mechanical aspects of the demon \cite{paper7,paper8,paper9,paper10,paper11,paper12,paper13,paper14,paper15}, and general relationships between the entropy and action of the demon from a  quantum information-theoretic point of view \cite{paper14,paper15}. On the other hand, the relationship between the work (or heat) and action of the demon is not yet fully understood  from this viewpoint. We stress that  $\Delta S = Q/ T$ is not valid in a general thermodynamic process.

Jarzynski has proved an irreversible-thermodynamic equality which relates the work to the free energy difference in an arbitrary isothermal process \cite{paper16,paper17}: $\langle \exp (- \beta W) \rangle = \exp (- \beta \Delta F)$, where $\beta = ( k_{\rm B} T )^{-1}$, $W$ is the work done on the system, $\Delta F$ is the difference in the Helmholtz free energy between the initial and final states, and $\langle \cdots \rangle$ is the statistical average over all microscopic paths. Note that this equality is satisfied even  when the external parameters are changed  at a finite rate. It follows from this equality that the fundamental inequality
\begin{eqnarray}
\langle W \rangle \geq \Delta F
\label{1}
\end{eqnarray}
holds.  While the original Jarzynski equality is classical, quantum-mechanical versions of the Jarzynski equality have been studied \cite{paper18,paper19,paper20,paper21}. Kim and Qian have recently generalized the equality for a classical Langevin system which is continuously controlled by a Maxwell's demon \cite{paper22}.

In this paper, we establish a general relationship between the work performed on a thermodynamic system and the amount of information gained from it by the demon, and prove the relevant equality and several corollary inequalities which are generalizations of Eq.~(\ref{1}). With the present setup, the demon performs a quantum measurement \cite{paper23,paper24} during an isothermal process, selects a sub-ensemble according to the outcome of the measurement, and performs unitary transformations on the system depending on the outcome. We follow the method of Ref.~\cite{paper14,paper24} to characterize the demon only in terms of its action on the system and do not make any assumption about the state of the demon itself. The subsequent results therefore hold true regardless of the state of the demon, be it in  equilibrium or out of equilibrium.

This paper is constituted as follows. In Sec.~I\hspace{-.1em}I, we formulate a general setup of isothermal processes with Maxwell's demon and illustrate the case of a generalized Szilard engine.  In Sec.~I\hspace{-.1em}I\hspace{-.1em}I, we derive the generalized Jarzynski equality, and new thermodynamic equalities generalizing inequality~(\ref{1}).  In Sec.~I\hspace{-.1em}V A,  we clarify the property of an effective information content obtained by the demon's measurement.  In Sec.~I\hspace{-.1em}V B, we discuss a crucial assumption of the final state of thermodynamic processes, which sheds light on a fundamental aspect of  the characterization of thermodynamic equilibrium states. Finally, in Sec.~V\hspace{-.1em}I\hspace{-.1em}I, we conclude this paper.

\section{Setup}

We consider an isothermal process at temperature $T=(k_{\rm B} \beta)^{-1}$, in which a thermodynamic system is in contact with an environment at the same temperature, and in which the initial and final states of the entire system are in thermodynamic equilibrium. We do not, however, assume that the states in the course of the process are in  thermodynamic equilibrium. We treat the isothermal process as the evolution of thermodynamic system S and  sufficiently large heat bath B, which are as a whole isolated and only come into contact with  some external mechanical systems and a demon. Apart from the  demon, the total Hamiltonian can be written as
\begin{eqnarray}
H^{\rm S+B}(t) = H^{\rm S} (t) + H^{\rm int} (t) + H^{\rm B},
\label{2}
\end{eqnarray}
where the time dependence of $H^{\rm S}(t)$ describes the mechanical operation on S through certain external parameters, such as an applied magnetic field or  volume of the gas, and the time dependence of $H^{{\rm int}} (t)$  describes, for example, the attachment of an adiabatic wall to S. We consider a time evolution from  $t_{\rm i}$ to $t_{\rm f}$, assume $H^{{\rm int}} (t_{\rm i}) = H^{{\rm int}} (t_{\rm f}) = 0$, and write $H^{\rm S+B} (t_{\rm i} ) = H_{\rm i}$ and $H^{\rm S+B} (t_{\rm f} ) = H_{\rm f}$. We consider the simplest isothermal process in the presence of the demon. This process can be divides into the following five stages:

\textit{Stage 1.}---At time $t_{\rm i}$, the initial state of S+B is in thermal equilibrium at temperature $T$. The density operator of the entire state is given by
\begin{eqnarray}
\rho_{\rm i} = \frac{\exp (- \beta H_{\rm i})}{Z_{\rm i}}, \ Z_{\rm i} = {\rm tr} \{ \exp (- \beta H_{\rm i}) \}. 
\label{3}
\end{eqnarray}
Note that the partition function of S+B is the product of that of S and that of B: $Z_{\rm i} = Z_{\rm i}^{\rm S}Z_{\rm i}^{\rm B}$, and the Helmholtz free energy of S+B is the sum $F_{\rm i} = F_{\rm i}^{\rm S} + F_{\rm i}^{\rm B}$, where $F_{\rm i} = - k_{\rm B} T \ln Z_{\rm i}$, etc.

\textit{Stage 2.}---From $t_{\rm i}$ to $t_1$, system S+B evolves according to  the unitary transformation represented by 
\begin{eqnarray}
U_{\rm i} =  T \exp \left( \frac{1}{i \hbar} \int_{t_{\rm i}}^{t_1} H^{\rm S+B} (t) dt \right).
\label{4}
\end{eqnarray}

\textit{Stage 3.}---From $t_1$ to $t_2$, a demon performs a quantum measurement described by measurement operators $\{ M_k \}$ on S and obtains each outcome $k$ with probability $p_k$. Let $K$ be the set of all the outcomes $k$ satisfying $p_k \neq 0$. Suppose that the number of elements in $K$ is finite. The process  proceeds to \textit{stage 4} only if  the outcome belongs to subset $K'$ of $K$, otherwise the demon discards the sample and the process restarts from \textit{stage 1}; we calculate the statistical average over subensemble $K'$. 

\textit{Stage 4.}---From $t_2$ to $t_3$, the demon performs a mechanical operation on S  depending on outcome $k$. Let  $U_k$ be  the corresponding unitary operator on S+B. We assume that the state of the system becomes independent of $k$ at the end of this stage, this being a feature characterizing the action of the demon \cite{paper14}. This stage describes a feedback control by the demon. Note that the action of the demon is characterized only by set $\{ K', M_k, U_k \}$.

\textit{Stage 5.}---From $t_3$ to $t_{\rm f}$, S+B evolves according to unitary operator $U_{\rm f}$ which is independent of outcome $k$. We assume that S+B has reached equilibrium at temperature $T$  by $t_{\rm f}$ from the macroscopic point of view because the degrees of freedom of B is assumed to far exceed that of S. The partition function of the final Hamiltonian is then given by
\begin{eqnarray}
\ Z_{\rm f} = {\rm tr} \{ \exp (- \beta H_{\rm f}) \}.
\label{5}
\end{eqnarray}
As in \textit{stage 1}, it follows that $Z_{\rm f} = Z_{\rm f}^{\rm S}Z_{\rm f}^{\rm B}$ and $F_{\rm f} = F_{\rm f}^{\rm S} + F_{\rm f}^{\rm B}$, where $F_{\rm i}^{\rm B} = F_{\rm f}^{\rm B}$. We denote as $\rho_{\rm f}$ the density operator of the final state.

After going through all the stages, the state of the system changes from $\rho_{\rm i}$ to 
\begin{eqnarray}
\rho_{\rm f} = \frac{1}{p} \sum_{k \in K'} E_k \rho_{\rm i} E_k^{\dagger},
\label{6}
\end{eqnarray}
where $E_k$ is given by
\begin{eqnarray}
E_k = U_{\rm f} U_k M_k U_{\rm i},
\label{7}
\end{eqnarray}
and $p$ is the sum of the probabilities of $k$'s belonging to $K'$ as
\begin{eqnarray}
p = \sum_{k \in K'} {\rm tr} (E_k^{\dagger}E_k \rho_{\rm i}).
\label{8}
\end{eqnarray}
Note that $E_k$ is a nonunitary operator because it involves the action of measurement by the demon. In contrast, in the original setup of Jarzynski, there is no demon and hence $E_k$ is unitary. 

In the case of a generalized Szilard engine, the foregoing general process can be illustrated as follows. We use a box containing a single molecule localized in a region  that is sufficiently smaller than the box. This molecule interacts with a  heat bath at temperature $T$. 

\begin{figure}
\includegraphics[width=0.95\linewidth]{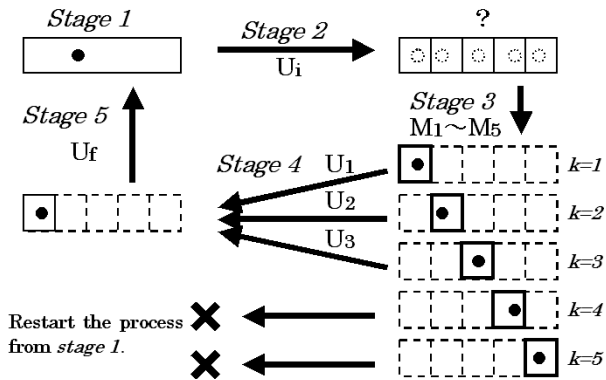}
\caption{A generalized Szilard engine with $n=5$ and $m=3$. The conventional Szilard engine corresponds to $n=m=2$. \textit{Stage 1.}---The state of the engine and the heat bath is initially in thermodynamic equilibrium. \textit{Stage 2.}---Divide the box into 5 partitions. \textit{Stage 3.}---A demon measures which partition the molecule is in. This measurement is described by measurement operators $\{ M_1, M_2, \cdots, M_5 \}$. \textit{Stage 4.}---The demon performs operation $U_k$ depending on measurement outcome $k$. If $k=1$, $2$ or $3$, the demon moves the $k$th box to the leftmost position. If the outcome is 4 or 5, the demon discards the sample and the process restarts from \textit{stage 1}. \textit{Stage 5.}---The box is expanded quasi-statically and isothermally so that the final state of the entire system returns to the initial state from a macroscopic point of view. This process is a unitary evolution of the entire system.  See the text for details.}
\label{figure1}
\end{figure}

\textit{Stage 1.}---A molecule is initially in equilibrium at temperature $T$. Let $\rho_{\rm i}$ be  the density operator of the initial state of the engine and the heat bath.

\textit{Stage 2.}---We divide the  box into $n$ partitions of equal volume. The state of the molecule then becomes $\rho_1 = (\rho (1) + \rho (2) + \cdots + \rho (n) )/n$, where $\rho (k)$ represents the state of the molecule in the $k$th partition. We set $K= \{ 1,2, \cdots , n \}$.

\textit{Stage 3.}---A demon performs a measurement on the system to find out where the molecule is. The demon chooses subset $K' = \{ 1,2, \cdots , m \}$ ($m \leq n$), and the process proceeds to \textit{stage 4} only if outcome $k$ belongs to $K'$. The state of the system is $\rho_2= (\rho (1) + \rho (2) + \cdots + \rho (m) )/m$ at the end of this stage.

\textit{Stage 4.}---When the outcome is $k$ ($\in K'$), the demon removes all but the $k$th box, and then moves the $k$th box to the leftmost position. This operation is described by  $U_k$. The state of the molecule after this operation is $\rho_3 = \rho (1)$.

\textit{Stage 5.}---We expand the box quasi-statically and isothermally so that the final state of the entire system returns to the initial state from a macroscopic point of view. Here by the last sentence we mean that the expectation value of any macroscopic quantity in the final state is the same as that in the initial state, as will be discussed in detail in Sec.~I\hspace{-1.em}V B.  This process is unitary in respect of the molecule and heat bath; we can thus describe this process by unitary operator $U_{\rm f}$. 

Figure \ref{figure1} illustrates these processes for the case of $n=5$ and $m=3$. When $n=m=2$, this process is equivalent to the conventional Szilard engine.

\section{Equality and inequalities}

Let us now prove the equality which constitutes the main result of this paper. Let $W$ be the work performed on S+B during the entire process, $\{ E_a^{\rm i} \}$ and $\{ | \varphi_a \rangle \}$ be the respective eigenvalues and eigenvectors of $H_{\rm i}$, and $\{ E^{\rm f}_b \}$ and $\{ | \psi_b \rangle \}$ the respective eigenvalues and eigenvectors of $H_{\rm f}$. We can then calculate the statistical average \cite{paper18,paper20} of $\exp (- \beta W)$ over the subensemble specified by condition $k \in K'$ as
\begin{eqnarray}
&{}& \langle \exp (- \beta W) \rangle_{K'} \nonumber \\
&=& \frac{1}{p} \sum_{a,b,k \in K'}   \frac{e^{- \beta E_a^{\rm i}}}{Z_{\rm i}} | \langle \psi_b | E_k | \varphi_a \rangle |^2 e^{- \beta (E^{\rm f}_b - E^{\rm i}_a)} \nonumber \\
&=& \frac{1}{p} \sum_{b,k \in K'} \langle \psi_b | E_k E_k^{\dagger} | \psi_b \rangle \frac{e^{- \beta E^{\rm f}_b}}{Z_{\rm i}} \nonumber \\
&=& \frac{Z_{\rm f}}{Z_{\rm i}} \frac{1}{p} \sum_{b,k \in K'} \langle \psi_b | U_{\rm f} U_k M_k M_k^{\dagger} U_k^{\dagger} U_{\rm f}^{\dagger} | \psi_b \rangle \frac{e^{-\beta E^{\rm f}_b}}{Z_{\rm f}},
\label{9}
\end{eqnarray}
where $p$ is given by Eq.~(\ref{8}). Making the polar decomposition of $M_k$ as $M_k = V_k \sqrt{D_k}$, $\ D_k = M_k^{\dagger}M_k$, where $V_k$ is a unitary operator, we can rewrite Eq.~(\ref{9}) as
\begin{eqnarray}
\langle  \exp (- \beta W)  \rangle_{K'} \! = \!  \frac{Z_{\rm f}}{Z_{\rm i}} \! \frac{1}{p} \! \sum_{k \in K'} \! {\rm tr}  ( \! D_k \! V_k^{\dagger} \! U_k^{\dagger} \! U_{\rm f}^{\dagger} \! \rho_{\rm f}^{\rm can} \! U_{\rm f} \! U_k \! V_k \! ),
\label{10}
\end{eqnarray}
where $\rho_{\rm f}^{\rm can}$ is the density operator of the canonical distribution of the final Hamiltonian:
\begin{eqnarray}
\rho_{\rm f}^{\rm can} = \frac{\exp (- \beta H_{\rm f})}{Z_{\rm f}}.
\end{eqnarray}
Let $\rho_1$ be the density operator just before the measurement, and $\rho_2^{(k)}$ be that immediately after the measurement with outcome $k$. We obtain $\rho_2^{(k)} = M_k \rho_1 M_k^{\dagger} / {\rm tr} (D_k \rho_1)$ and $\rho_2^{(k)} = U_k^{\dagger} U_{\rm f}^{\dagger} \rho_{\rm f} U_{\rm f} U_k$; therefore
\begin{eqnarray}
V_k^{\dagger} U_k^{\dagger} U_{\rm f}^{\dagger} \rho_{\rm f} U_{\rm f} U_k V_k = \frac{\sqrt{D_k} \rho_1 \sqrt{D_k}}{{\rm tr} (D_k \rho_1)}.
\label{11}
\end{eqnarray}
Thus Eq.~(\ref{10}) reduces to 
\begin{eqnarray}
\langle \exp (- \beta W) \rangle_{K'} =  \frac{Z_{\rm f}}{Z_{\rm i}} \frac{\eta}{p} \left( 1+ \frac{\Delta \eta }{\eta} \right),
\label{12}
\end{eqnarray}
where we introduced the notation
\begin{eqnarray}
\eta \equiv \sum_{k \in K'} \frac{{\rm tr} (D_k^2 \rho_1)}{{\rm tr} (D_k \rho_1)} = \sum_{k \in K'} \frac{{\rm tr} ((E_k^{\dagger}E_k)^2 \rho_{\rm i})}{{\rm tr} (E_k^{\dagger}E_k \rho_{\rm i})}, \\
\Delta \eta \equiv \sum_{k \in K'} {\rm tr} (D_k V_k^{\dagger} U_k^{\dagger} U_{\rm f}^{\dagger} (\rho_{\rm f} - \rho_{\rm f}^{\rm can}) U_{\rm f} U_k V_k).
\label{13}
\end{eqnarray}
The parameter $\eta$ describes the measurement error of the demon's measurement, the precise meaning of which is discussed in Sec. V.
We now \textit{assume} that the final state $\rho_{\rm f}$ satisfies
\begin{eqnarray}
\Delta \eta = 0.
\label{assumption}
\end{eqnarray}
We discuss in detail the physical meaning and validity of this assumption in Sec. I\hspace{-.1em}V B. Note that if the density operator of the final state is the canonical distribution, i.e. $\rho_{\rm f} = \rho_{\rm f}^{\rm can}$, then the above assumption~(\ref{assumption}) is  trivially satisfied.  Under this assumption, we finally obtain 
\begin{eqnarray}
\langle \exp (- \beta W) \rangle_{K'} = \exp \left( - \beta \Delta F + \ln \frac{\eta}{p} \right),
\label{14}
\end{eqnarray}
where $\Delta F = F_{\rm f} - F_{\rm i} = F^{\rm S}_{\rm f} - F^{\rm S}_{\rm i}$. This is the main result of this paper.

In a special case in which $D_k$'s are projection operators for all $k$, and $K' =K$, Eq.~(\ref{14}) reduces to 
\begin{eqnarray}
\langle \exp (- \beta W) \rangle = \exp \left( - \beta \Delta F + \ln d \right),
\label{17}
\end{eqnarray}
where $d$ is the number of elements in $K$. Note that the right-hand side of Eq.~(\ref{17}) is independent of the details of  pre-measurement state $\rho_1$.

We can apply the generalized Jarzynski equality to prove an inequality. It follows from the concavity of the exponential function that
\begin{eqnarray}
\exp (- \beta \langle W \rangle_{K'}) \leq \langle \exp (- \beta W) \rangle_{K'};
\label{15}
\end{eqnarray}
we therefore obtain
\begin{eqnarray}
\langle W \rangle_{K'} \geq \Delta F - k_{\rm B} T \ln \frac{\eta}{p}.
\label{16}
\end{eqnarray}

Consider the case in which the demon selects single outcome $k$, that is $K' = \{ k \}$.  Equation~(\ref{9}) then reduces to $\langle \exp (- \beta W) \rangle_k = \exp ( - \beta \Delta F + (\eta_k / p_k))$, where $\eta_k = {\rm tr} (D_k^2 \rho_1) / {\rm tr} (D_k \rho_1)$, and inequality (\ref{16}) becomes 
\begin{eqnarray}
\langle W \rangle_{k} \geq \Delta F - k_{\rm B} T \ln \frac{\eta_k}{p_k}.
\label{18}
\end{eqnarray}
Averaging inequality (\ref{18}) over all $k \in K$, we obtain
\begin{eqnarray}
\langle W \rangle \geq \Delta F - k_{\rm B} T H^{\rm eff},
\label{19}
\end{eqnarray} 
where
\begin{eqnarray}
H^{\rm eff} \equiv \sum_{k \in K} p_k \ln \frac{\eta_k}{p_k},
\label{20}
\end{eqnarray}
describes an effective information content which the demon gains about the system. Inequality (\ref{19}) is a generalization of (\ref{1}) and is stronger than (\ref{16}). It shows that we can extract work  larger than $- \Delta F$ from a single heat bath in the presence of the demon, but that we \textit{cannot} extract work larger than $k_{\rm B} T H^{\rm eff} - \Delta F$.

\section{Discussions}

\subsection{Effective Information Content}

We discuss the physical meaning of $\eta_k$ and $H^{\rm eff}$. It can  easily be shown that
\begin{eqnarray}
p_k \leq \eta_k \leq 1.
\label{21}
\end{eqnarray}
Here $p_k = \eta_k$ for all $\rho_1$ if and only if $D_k$ is proportional to the identity operator, and $\eta_k = 1$ for all $\rho_1$ if and only if  $D_k$ is a projection operator. In the former case,  the demon can gain no information about the system, while in the latter case, the measurement is error-free. Let us consider the case of $D_k = P_k + \varepsilon P_l$ ($l \neq k$), where $P_k$ and $P_l$ are projection operators and $\varepsilon$ is a small positive number. Then $\eta_k$ is given by
\begin{eqnarray}
\eta_k = \frac{{\rm tr} (D_k^2 \rho_1)}{{\rm tr} (D_k \rho_1)} = 1- \varepsilon \frac{{\rm tr} (P_l \rho_1)}{{\rm tr} (P_k \rho_1)} + o (\varepsilon).
\label{22}
\end{eqnarray}
We can therefore  say that  $1-\eta_k$ is a measure of distance between $D_k$ and the projection operator.  It follows from (\ref{21}) that 
\begin{eqnarray}
0 \leq H^{\rm eff} \leq H,
\label{23}
\end{eqnarray}
where $H$ is the Shannon information content that the demon obtains: $H = - \sum_{k \in K} p_k \ln p_k$.

We now derive some special versions of inequality (\ref{19}).  If the demon does not get information (i.e., $H^{\rm eff} =0$), inequality (\ref{19}) becomes $\langle W \rangle \geq \Delta F$, which is simply inequality (\ref{1}). On the other hand, in the case of a projection measurement, where $H^{\rm eff} =H$, (\ref{19})  becomes $\langle W \rangle \geq \Delta F - k_{\rm B} T H$.  An inequality similar (but not equivalent) to this inequality has been proved by Kim and Qian for a classical Langevin system \cite{paper22}.


\subsection{Characterization of Thermodynamic Equilibrium States}

We next show the physical validity of the assumption~(\ref{assumption}). 

In general, the canonical distribution describes the properties of thermodynamic equilibrium states.  However, the thermodynamic equilibrium is a macroscopic property characterized only by the expectation values of macroscopic quantities. In fact, to show that a heat bath is in thermodynamic equilibrium, we do not observe each molecule, but observe only macroscopic quantities of the heat bath, for example the center of mass of a ``small''  fraction involving a large number (e.g. $\sim 10^{12}$) of molecules.  Thus a density operator corresponding to a thermodynamic equilibrium state is \textit{not} necessarily the rigorous canonical distribution; $\rho_{\rm f} = \rho_{\rm f}^{\rm can}$ is too strong an assumption.

Note that no assumption has been made on the final state $\rho_{\rm f}$ in the derivation of the original Jarzynski equality without Maxwell's demon~\cite{paper16}, so $\langle \exp (- \beta W) \rangle = \exp (- \beta \Delta F)$ holds for any final state $\rho_{\rm f}$.  Under the condition that the final state of S+B is in thermodynamic equilibrium corresponding to the free energy $F_{\rm f}$ from a macroscopic point of view, we can interpret inequality~(\ref{1}), which can be shown from the Jarzynski equality, as the thermodynamic inequality for a transition between two thermodynamic equilibrium states, even if $\rho_{\rm f}$ is not a rigorous canonical distribution.

On the other hand, we have required the supplementary assumption~(\ref{assumption}) for the final state $\rho_{\rm f}$ to prove the generalized Jarzynski equality with Maxwell's demon.  The assumption~(\ref{assumption}) holds if
\begin{eqnarray}
{\rm tr} ( \tilde{D}_k \rho_{\rm f} ) = {\rm tr} (\tilde{D}_k \rho_{\rm f}^{\rm can}),
\label{a}
\end{eqnarray}
for all $k$ in $K'$, where $\tilde{D}_k \equiv U_{\rm f}  U_k  V_k D_k  V_k^{\dagger}   U_k^{\dagger}  U_{\rm f}^{\dagger}$.  We can say that under the assumption~(\ref{assumption}), $\rho_{\rm f}$ is restricted not only in terms of macroscopic quantities, but also constrained so as to meet Eq.~(\ref{a}).  It appears that the latter constraint is connected with the fact that the system in state $\rho_{\rm f}$ and that in state $\rho_{\rm f}^{\rm can}$ should not be distinguished by the demon.

We stress that our assumption~(\ref{assumption}) is extremely weak compared with the assumption $\rho_{\rm f} = \rho_{\rm f}^{\rm can}$. To see this, we denote as $f$  the degree of freedom of S+B (e.g. $f \sim 10^{23}$), as $N$ the dimension of the Hilbert space corresponding to S+B, and as $d'$ the number of elements in $K'$.  We can easily show that $N \geq O(2^f)$. On the other hand, $d' = O(1)$ holds in  a situation that the role of the demon is experimentally realizable. For example, suppose that S is a spin-1 system and B consists of harmonic oscillators, and the demon made up from optical devices performs a projection measurement on S.  In this case, $d' \leq d=3$ and $N=\infty$.  The rigorous equality $\rho_{\rm f} = \rho_{\rm f}^{\rm can}$ holds, in general, if and only if  all the matrix elements in $\rho_{\rm f}$ coincides with that of $\rho_{\rm f}^{\rm can}$, and note that the number of independent real valuables in the density operators is $N^2-1$.  On the other hand, only $d'$ equalities in~(\ref{a}) are required to meet the assumption $\Delta \eta = 0$.  

Although it is a conjecture that the assumption~(\ref{assumption}) is virtually realized in real physical situations, we believe that it is indeed realized in many situations.  A more detailed analysis on the assumption~(\ref{assumption}) is needed to understand the concept of thermodynamic equilibrium.

Finally, we consider the case that the assumption~({\ref{assumption}}) is not satisfied and 
\begin{eqnarray}
\Bigl|{\rm tr} \left( \tilde{D}_k ( \rho_{\rm f}  -   \rho_{\rm f}^{\rm can} )  \right) \Bigr| \leq \varepsilon
\label{b}
\end{eqnarray}
holds for all $k$ in $K'$. We can then estimate the value of $|  \Delta \eta / \eta |$ as
\begin{eqnarray}
\Bigl|\frac{\Delta \eta}{\eta} \Bigr| \leq \frac{d'}{p} \varepsilon.
\end{eqnarray}
Thus the deviation from the generalized Jarzynski equation is bounded only by the difference on the left-hand side of Eq.~(\ref{b}).

\section{Conclusion}

In conclusion, we have generalized the Jarzynski equality to a situation involving Maxwell's demon and derived several inequalities in thermodynamic systems. The demon in our formulation performs a quantum measurement and a unitary transformation depending on the outcome of the measurement. Independent of the state of the demon, however, our equality~(\ref{17}) establishes a close connection between the work and the information which can be extracted from a thermodynamic system, and our inequality~(\ref{19}) shows that one can extract from a single heat bath work greater than $- \Delta F$ due to an effective information content that the demon gains about the system.  To analyze broader aspects of information thermodynamic processes merits further study.

\begin{acknowledgments}
This work was supported by a Grant-in-Aid for Scientific Research (Grant No.\ 17071005) and by a 21st Century COE program at Tokyo Tech, ``Nanometer-Scale Quantum Physics'', from the Ministry of Education, Culture, Sports, Science and Technology of Japan. MU acknowledges support by a CREST program of JST.
\end{acknowledgments}


\end{document}